\documentclass[a4paper,20pt]{article}
    \usepackage[T1]{fontenc}
    \usepackage{graphicx}
    \usepackage{epsfig}
    \usepackage{amsmath}
    \usepackage{amsfonts}
    \renewcommand{\abstract}{}
\begin{document}
\makeatletter
\renewcommand{\@oddhead}{\textit{YSC'14 Proceedings of Contributed Papers} \hfil \textit{K. Kowalik, M. Hanasz}}
\renewcommand{\@evenfoot}{\hfil \thepage \hfil}
\renewcommand{\@oddfoot}{\hfil \thepage \hfil}
\fontsize{11}{11} \selectfont

\title{Initial Magnetization of Galaxies by Exploding, Magnetized Stars}
\author{\textsl{K. Kowalik$^1$, M. Hanasz$^1$}}
\date{}
\maketitle
\begin{center} {\small $^{1}$Torun Centre for Astronomy of Nicolaus Copernicus University \\
Kacper.Kowalik@astri.umk.pl, Michal.Hanasz@astri.umk.pl}
\end{center}

\begin{abstract}
We conduct a series of magnetohydrodynamical (MHD) simulations of magnetized
interstellar medium (ISM) disturbed by exploding stars. Each star deposits a
randomly oriented, dipolar magnetic field into ISM. The simulations are
performed in a Cartesian box, in a reference frame that is corotating with the
galactic disk. The medium is stratified by vertical galactic gravity. The
resulting turbulent state of ISM magnetized by the stellar explosions is
processed with the aid of Fourier analysis. The results leads to the conclusion
that the input of magnetic energy from exploding stars is additionally multiplied
by differential rotation. The resulting  magnetic field appears to grow up in
small-scale component, while the total magnetic flux remains limited.
Our results indicate that magnetic field originating from exploding stars can
be a source of initial magnetic fields for a subsequent dynamo process.
\end{abstract}

\section*{Introduction}

\indent \indent There is a strong observational evidence that
magnetic fields are present in virtually all galaxies. It is
commonly believed that those fields are generated due to an
$\alpha\omega$ dynamo process, where differential rotation
$(\omega)$ and helical turbulence $(\alpha)$ are responsible for
creating a strong, large--scale magnetic field from a weak,
small--scale initial one \cite{parker}. The dynamo can amplify and
restructure the magnetic field (see i.e. \cite{widrow} for a recent
review of galactic dynamo theory), yet it cannot create a new one,
thus a seed field is required. Although, the origin of the seed,
magnetic field is a mystery yet to be solved, a few theories
concerning the problem exist. One of these theories \cite{rees}
points to the very first generation of stars as a possible source of
the seed, magnetic fields. \par Even if a star is born without any
primordial magnetic field, any nonparallelism between the gradient
of pressure and the gradients of thermodynamical quantities like
density or temperature, results in non--vanishing
$\nabla\times\mathbf{E}$ (more details in \cite{kemp}). That,
according to Faraday's law, implies time dependent magnetic field.
This effect is known as Biermann battery process \cite{biermann}.
Eventually, the newly created magnetic field is amplified by a
stellar dynamo. If, during its evolution, the star explodes as a
supernova or undergoes a significant mass loss, the
,,frozen--in--plasma'' magnetic field is spread throughout the ISM,
initiating the $\alpha\omega$ dynamo. The aim of this paper is to
verify experimentally the hypothesis presented by Rees \cite{rees}
that young galaxies have been magnetized by processes of stellar
origin. As suggested in \cite{rees}, if we consider that supernova
remnant (like the Crab Nebula) deposits flux of order $10^{34}$ G
cm$^2$, then $N$ such remnants would increase the net flux in galaxy
by a factor $N^x$, where $x\in[1/3,1/2]$. As far as the authors
know, nobody has ever tried to test this hypothesis in a numerical
experiment (however, a paper concerning quite similar problem was
recently published \cite{ziegler}). Although this quantitative
estimation seems to be confirmed by the results of this paper, some
qualitatively new effects are being found.

\section*{Physical setup and numerical model}

\indent \indent We assume that gas forming galactic disk is
completely ionized, and apply the standard set of MHD equations (see
\cite{jackson}), supplemented with the vertical gravitational
acceleration and rotational pseudo-forces in the equation of gas
motion
\begin{align}
\label{mhd1}
  \frac{\partial\mathbf{v}}{\partial t} + (\mathbf{v}\cdot\nabla)\mathbf{v} &=
      - \frac{1}{\varrho}\nabla \left( p+ \frac{B^2}{8\pi} \right) - 2\mathbf{\Omega}\times\mathbf{v} + \Omega^2 x \hat{x} - g_z
   + \frac{(\mathbf{B}\cdot \nabla)\mathbf{B}}{4\pi\varrho}, \\
   \frac{\partial\varrho}{\partial t} + \nabla\cdot (\varrho\mathbf{v}) &= 0, \\
   \varrho \left( \frac{\partial\epsilon}{\partial t} +
  \mathbf{v}\cdot\nabla \epsilon \right) + p\nabla\mathbf{v} &= 0,\\
  \frac{\partial\mathbf{B}}{\partial t} &= \nabla \times(\mathbf{v}\times\mathbf{B}), \label{ind-eq} \\
  \nabla\cdot\mathbf{B}&=0,\\
  \mathbf{j}&=\frac{c}{4\pi}\nabla\times\mathbf{B},\label{mhd2}
\end{align}
with addition of the adiabatic equation of state
\begin{equation} p = (\gamma -1)\varrho\epsilon \end{equation}
with $(\gamma = 5/3)$.

To solve the set of partial differential equations numerically we apply our own
parallelized 3D MHD code based on the {\em relaxing TVD}  scheme \cite{jx},
which is described in details by Trac and Pen \cite{tvd}
and extended for MHD system of equations by Pen et al. \cite{tvd2}. The
algorithm of magnetic field evolution, based on the constraint transport (CT)
algorithm \cite{eh}, preserves the divergence-free magnetic field
at the machine accuracy.

We chose a reference frame corotating with the disk, at the
$R_0=R_\odot$ (where $R_\odot$ is Sun's galactic radius) and use, in
addition to rotational pseudo-forces the shearing-periodic boundary
conditions \cite{hawley}, which are a modification of periodic
boundary conditions, that is designed to model differentially
rotating astrophysical disks. For further details concerning
shearing box see Gressel and Ziegler \cite{Gres2}. We introduce the
local reference frame by adding the terms of Coriolis force
$2\mathbf{\Omega}\times\mathbf{v}$ and the tidal expansion of the
combined, effective centrifugal and gravitational potential about
$R_0$ --- $\Omega^2x$, to the equation of motion (\ref{mhd1}).
Following Ferriere \cite{katia} the vertical component of galactic,
gravitational acceleration ($g_z$ term in (\ref{mhd1})) can be
expressed as

\begin{equation}
-g_z(R_\odot,z) = (4.4\cdot10^{-9}\textrm{ cm s}^{-2})\frac{z}{\sqrt{z^2+(0.2\textrm{ kpc})^2}} + (1.7\cdot10^{-9}\textrm{ cm s}^{-2})\frac{z}{1\textrm kpc}
\end{equation}
\section*{Numerical simulations}

\indent \indent We perform numerical simulations of the interstellar
medium described above, perturbed with randomly distributed
magnetized supernova (SN) explosions. Each stellar explosion
deposits  the dipolar magnetic field within a spherical region of
radius 10 pc. The computational domain represents a rectangular
region of $0.5 {\rm kpc} \times0.5{\rm kpc}\times1.5{\rm kpc}$ in
$x$, $y$ and $z$ directions respectively, and the grid resolution is
$125\times125\times375$ cells. We assume that stellar explosions are
uniformly distributed across the galactic plane, whereas vertical
distribution is normal,  with $\sigma=100$ pc. In the present local
approximation we neglect the effect of spiral arms, since our
computational domain covers only a small volume of the galactic
disks. In this approach one could consider a time modulation of the
supernova rate, corresponding to the passages of spiral arms through
the computational volume, however, we do not expect a significant
effects of this modulation on long timescales. Each explosion is
realized by adding thermal energy to the gas in sphere of radius
$10$ pc. The explosion energy is scaled down by several orders of
magnitude with respect to the real SN energy output, due to
limitations of the present version of our code. Furthermore, each
explosion deposits a randomly oriented (directions distributed
uniformly on sphere) dipolar magnetic field $\mathbf{B}_{\rm dip} =
\nabla\times\mathbf{A}$, where
\begin{equation}
\mathbf{A}(r,\varphi,\theta) = A_0 \frac{r\sin\theta}{(l^2 + r^2 +2rl\sin\theta)^{3/2}}\mathbf{e}_\varphi
\end{equation}
where $\mathbf{A}(r,\varphi,\theta)$ is a vector potential of a dipolar
magnetic  field created by an electric current in toroidal circuit of final diameter $l$ \cite{jackson}, corresponding to the above mentionned size of
SN remnant. The assumed density in the galactic plane is $\varrho_0
= 0.32564 \textrm{M}_\odot/\textrm{pc}^3\sim 13 \textrm{ atom/cm}^3$ and star
explosions rate is $\sigma = 20 \textrm{ kpc}^{-2}\textrm{ Myr}^{-1}$. Both
quantities $\rho_0,\,\sigma$ are derived from recent observational data
\cite{katia}.

\section*{Results}

\indent \indent In this section we discuss the evolution of the
interstellar medium which is subject to a gradual magnetization by
exploding stars.  A typical snapshots displaying greyscale--coded
gas density and magnetic vectors $\mathbf{B}$ at $t=30$ Myr are
shown  in Fig.1. According to expectations, magnetic field in the
disk volume displays a random configuration, which results as a
superposition of randomly oriented small-scale dipolar magnetic
fields. The fluctuations of gas density result from the input of
magnetic and thermal energy in each explosion region.

As we can see in Fig.2, the exponential growth of the mean magnetic
flux is visible during the first phase of the simulation. However,
after roughly $60$ Myr magnetic flux cease to grow, whereas magnetic
energy continues to grow due to the ongoing SN explosions activity.
We show in Fig.3 a plot of total magnetic energy scaled to the
supplied magnetic energy, and spectrum of magnetic energy
fluctuations, as a function of time.

As it is apparent in  Fig.3, the total magnetic energy grows faster
than one would expect from simple summation of magnetic energies
from individual explosion events. The growth of  magnetic energy is
apparently enhanced by differential rotation, which amplifies the
toroidal magnetic field component via stretching the radial magnetic
field. This effect is described by the induction equation
(\ref{ind-eq}), which implies the following approximated equation
for the azimuthal magnetic field
\begin{equation}
\frac{\partial B_\varphi}{\partial t} \simeq G B_r,  \label{omega}
\end{equation}
where $G =r d\Omega/dr \simeq \Omega$ is the measure of differential
rotation. Since the galactic angular velocity applied in our
simulation is  $\Omega=0.05$, the toroidal magnetic field is
generated on a timescale of 20 Myr. The respective growth time of
magnetic energy should be twice shorter. Since the dipolar magnetic
field, is supplied into the initially unmagnetized medium, the
growth of magnetic energy is initially slow, but later on the
amplification of magnetic field by differential rotation speeds up.
As it is apparent in Fig.3, the observed growth time of magnetic
energy is consistent with our estimation.

In the right panel of Fig.3 we show the spectrum of magnetic energy
at $t=25$ Myr and $t=119$ Myr, along $x$ and $y$ directions,
obtained by means of Fourier analysis. The two straight lines
corresponding to the slope -5/3, shown for comparison, represent the
Kolmogorov's spectrum. The spectral analysis of  magnetic energy in
two different time instants shows a week tendency of steepening of
the spectrum of magnetic field fluctuations along the $x$-direction
and flattening in $y$ direction.   The results presented in Fig.3
mean that the spectrum of magnetic fluctuations, which is strongly
anisotropic at the beginning of the experiment becomes more and more
isotropic in course of time. The overall spectrum of magnetic
fluctuations at the end of our simulation remains relatively flat
with respect to Kolmogorov's spectrum. Magnetic energy cumulated on
small spatial scales remains large in comparison to the energy on
large scales. Although the simulation period of 120 Myr is still
short with respect to the galactic rotation period of the order of
200 Myr, one can say that the evolution of magnetic spectrum is
rather weak.

\section*{Conclusions}

\indent \indent Our results indicate that the structure of stellar
origin galactic magnetic field does not fit to the current picture
of polarimetric radio-observations of disk galaxies (see
\cite{widrow} for references on observational results of galactic
magnetic fields). The resulting magnetic field configuration can
serve as an initial condition for further exploration of
$\alpha\omega$ dynamo process.

In the next step we plan to extend our physical setup with the
cosmic ray component, described by the diffusion-advection equation,
as it has been done by Hanasz and Lesch \cite{hnl}. The presence of
cosmic rays leads inevitably to the Parker instability, and a very
efficient $\alpha\omega$-dynamo process \cite{hanasz1}. We also plan
to  extend the time of the simulations to at least $1$ Gyr, and to
simulate the whole galactic disk in full 3D, instead of using local
approximation.  Summarising, note the following effects of random,
magnetized supernova explosions in the differentially rotating
interstellar medium:
\begin{itemize}
\item limited growth of magnetic flux accompanied with cumulation of energy in
small-scale magnetic fields;
\item an additional effect of magnetic field amplification by differential
rotation;
\item a relatively slow evolution  of magnetic  spectrum, indicating that a
subsequent dynamo process involving inverse turbulent cascade is necessary to
obtain results compatible with magnetic maps of real galaxies.
\end{itemize}

\section*{Acknowledgements}
\indent \indent This work was supported by the Ministry of Science
and Higher Education of Poland through the grant 1/P03D/004/26. The
presented computations have been performed on the HYDRA beowulf
cluster in Torun Centre for Astronomy.

\newpage

\textbf{Figure 1.} Slices through the computational domain showing
density magnetic field at $T = 30$ Myr. The left panel shows a
vertical slice through the domain at $y=0$ pc, while the right panel
represents a horizontal slice at $z=0$ pc.\vspace{10ex}

\textbf{Figure 2.} The total magnetic energy (left panel) and
evolution of the mean magnetic flux in time (right
panel).\vspace{10ex}

\textbf{Figure 3.} Temporal evolution of the total magnetic energy
scaled to magnetic energy supplied in supernova remnants (left
panel) and magnetic energy spectrum analyzed along $x$ and $y$
directions (right panel).\vspace{10ex}

Figures are available on YSC home page
(http://ysc.kiev.ua/abs/proc14$\_$11.pdf).

\end{document}